# Real-time Building Energy Storage Scheduling under Electrical Load Uncertainty: A Dynamic Markov Decision Process Approach with Comprehensive Analysis of Different Pricing Policies

Hussein Sharadga[1, *], Ahmad Dawahdeh[2], Golbon Zakeri[3], Abdullah Hayajneh[4], Geoff Pritchard[5]

[1]Texas A&M International University, [2]Jordan University of Scince and Technology, [3]University of Massachusetts Amherst, [4]Texas A&M University, [5]The University of Auckland, [*] Corresponding email: hssharadga@utexas.edu

## Abstract

In response to the increasing deployment of battery storage systems for cost reduction and grid stress mitigation, this study presents the development of a new real-time Markov decision process model to efficiently schedule battery systems in buildings under electrical load uncertainty. The proposed model incorporates quantile Fourier regression for load fitting, leading to a large-scale optimization problem with approximately a million decision variables and constraints. To address this complexity, the problem is formulated as a linear program and solved using a commercial solver, ensuring effective navigation and identification of optimal solutions. The model's performance is evaluated by considering different pricing policies and scenarios, including demand peak shaving. Validation of the Markov model is conducted using one year of historical demand data from a school. Findings indicate that MDP performance in adapting to uncertain loads can range from 30 to 99% depending on the pricing policy.

**Funding**

This work was supported by the United States National Science Foundation (NSF) under grants DGE-2021693 and SES-2020888.

**Competing Interests**

The authors declare that they have no known competing financial interests or personal relationships that could have appeared to influence the work reported in this paper.

## 1. Introduction

The electricity demand has been on the rise due to the increasing population and the growing electrification such as adoption of electric heating systems and transportation. This heightened demand places significant stress on the grid, necessitating costly upgrades to the infrastructure. To address this challenge, battery storage systems have been deployed to alleviate stress on the grid and prevent power outages by shifting the demand. However, scheduling battery systems remains a challenging optimization problem due to fluctuating weather conditions and varying electrical loads.

Many control algorithms have been developed to schedule battery systems under uncertainty. Li et al. [1] proposed a novel centralized strategy to shave the battery charging peak under load fluctuations. In [2], stochastic dual dynamic programming was used to dispatch battery systems under electrical load uncertainty. To achieve better operation, battery aging is considered in the decision-making, and the battery is scheduled with a receding horizon. Stochastic programming combined scenarios tree handled the load and solar photovoltaic uncertainties in [3]. Chance-constrained programming is used to schedule solar photovoltaic battery systems under solar energy uncertainty in [4]. The probabilistic chance-constraint formula is simplified into a deterministic form by fitting solar energy on a probability distribution. Der Meer et al. [5] utilized scenario-based programming to schedule solar photovoltaic battery systems to maximize energy self-consumption. To reduce the size of the problem, the load and the solar energy were replaced with net power. The conditional-value-at-risk was used in [6] to operate battery systems integrated with the power grid under varying electricity prices. In [7], solar photovoltaic battery systems were used to shave the demand peak. The system was scheduled by model predictive control. In another study [8], the model predictive control was utilized to achieve a balance between the system degradation and the self-consumption of energy.

Markov decision processes (MDPs) constitute another well-known model to control stochastic processes. Markov decision process applications are found in sales promotion, design of experiments, water resources, agriculture, and energy [9]. MDP has been applied to solve various problems in the energy sector. In [10], MDP was employed to optimize the supply bidding process. A novel optimization framework to minimize the cost of water distribution for water utilities was proposed in [11] based on MDP. In [12], MDP is utilized to keep the smart handhelds from being active for a long time to reduce energy waste. MDP energy-efficient irrigation model for agriculture applications was developed in [13]. MDP is utilized to schedule integrated hydrogen energy systems in [14].

MDP has great potential in scheduling battery storage systems. In [15], MDP is used to coordinate between two battery storages of different types: a lead acid battery system and a vanadium redox flow battery. MDP prioritizes the charging and discharging process based on the current conditions. A real-time dynamic MDP is used to operate a charging station supplied by a solar photovoltaic system and integrated with the grid energy under the (dynamic) France electricity prices in [16]. In [17], MDP was implemented to schedule energy storage systems in distribution networks with renewable energy. Markov chain modeled the variation in the renewable generation. MDP produces an optimal battery scheduling policy to minimize the cost of electricity and network losses. A power allocation smoothing strategy is developed in [18] based on MDP for hybrid energy storage systems to mitigate the fluctuations in the power without significantly increasing the computation time. Iversen et al. [19] proposed an algorithm to charge electric vehicles based on MDP. The MDP maps the vehicle's use, the electricity price, and the end-user's risk aversion to an optimized policy. The proposed study assumed a day-ahead electricity market. In [20], MDP was utilized to schedule electricity purchases and sales to the power grid. The model's inputs are storage's current state of charge, future electricity consumption, and generation.

The current body of research lacks comprehensive evaluations of MDP performance in real-time scheduling of battery systems, particularly when considering different pricing policies, such as time-of-use, real-time pricing, energy-limit penalty, and peak demand disincentives, as well as varying operating conditions that can occur throughout the year. Moreover, there is a notable absence of any previous proposals that leverage quantile Fourier regression to mitigate the uncertainty associated with MDP policies in the presence of load uncertainty. In this paper, a Markov decision process is proposed to construct an hourly schedule to dispatch battery systems under electrical load uncertainty to reduce the building electricity costs. The effectiveness of the MDP approach lies in its capability to optimize decisions considering long-term consequences, in the presence of uncertainty. By employing MDP, a policy is generated that remains valid for every time period throughout the year by adapting to the variations in the electrical load. This aspect is particularly advantageous as it creates a robust policy that does not necessitate periodic updates or adjustment. Our study is conducted for different pricing policies of varying complexity, with demand peak shaving being the most complex. The MDP is trained in this study by the load's probability of transition. The transition probability matrix of the load is extracted by fitting the load using quantile Fourier regression. Quantile Fourier regression is proposed to cut load's uncertainty. That, however, produces a large-scale optimization problem with approximately one million decision variables and a million constraints. The problem has been formulated into a linear program solved using a commercial solver. A year of historical data is used for training MDP and another year for testing. The design pipline of the proposed MDP policy is outlined in **Fig. 1**.

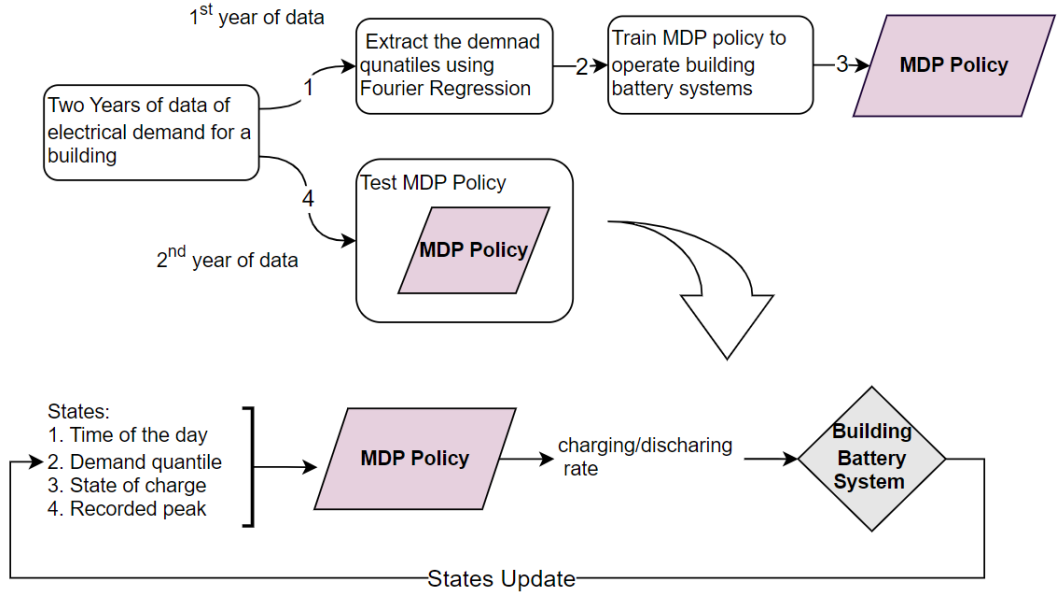

**Fig. 1** Flowchart describing the proposed MDP policy design pipeline

The paper is structured as follows: In Section 2, the detailed process of extracting load quantiles using Fourier regression is presented. Following that, we introduce the proposed battery scheduling model that is based on MDPs. Additionally, the MDP model is reformulated to solve the peak shaving problem in this section. Section 3 focuses on a comprehensive case study conducted to evaluate the effectiveness of the proposed method in real-time scheduling of battery systems under load uncertainty. The performance of the MDP approach is rigorously validated through a comparison with the deterministic case, taking into account different pricing policies. Finally, in Section 4, we conclude our study, summarizing the key findings.

**2. Problem Formulation**

In this section we lay out the problem formulation where we develop a real-time optimization framework based on MDPs to schedule battery systems under load uncertainty. One year of historical data is utilized to extract the demand quantiles using Fourier regression in **Section 2.1**. The demand quantiles over the historical year are used to generate the probability transition matrix for the Markov chain that estimates the electricity demand process in **Section 2.2**. The battery scheduling process is formulated as a Markov decision process in **Section 2.3**. The battery dispatching model and its constraints are given in **Section 2.4.** Subsequently, in the last section, we will formulate the Markov decision process model for demand peak shaving problem**.**

## 2.1. The Demand Quantiles

Electricity demand varies diurnally (and seasonaly). In particular, electricity demand has daily patterns of morning and evening peak, while midnight and early hours in the morning are likely to be periods of low usage. Hence, it is useful to employ periodic functions to capture the nature of electricity demand. We employ Fourier regressions (using trigonometric functions) [21] to fit our demand data. Rather than merely model demand for each period by its mean, i.e. use a regression to portray the trend, we employ Fourier regressions for different quantiles of the data. This approach will allow for a more accurate stochastic process model of demand that can lead to enhanced operational decisions. The demand quantiles are obtained by solving the following minimization problem [22]:

$$\text{Minimize } MAD = \frac{1}{N} \sum_{i=1}^{N} \rho_\beta (e_i) \tag{1}$$

where $MAD$ is the median absolute deviation, $N$ is the number of samples, and $\rho_\beta$ is a quantile-based function, given as follows:

$$\rho_\beta(e_i) = \beta \max(e_i, 0) + (1 - \beta) \max(-e_i, 0) \tag{2}$$

$e$ is the error, the deviation between the actual value and the value predicted by Fourier regression. $e$ is calculated as follows:

$$e_i = x_i - [\mu + A_n \cos(2n\pi f t) + B_n \cos(2n\pi f t)] \tag{3}$$

where $x_i$ is the actual value, $\mu$ is added constant, $A$ and $B$ represent the amplitudes of the sine and cosine waves, and $n$ is the number of Fourier waves. Here, $f$ is the frequency of the waves, $f = 1/T$. $T$ is the period of electrical load and is assumed to be one day (diurnal variation).

The convex nature of this optimization problem makes it suitable for solving using convex optimization techniques. Convex optimization offers fast convergence, which becomes particularly advantageous when dealing with a substantial number of Fourier terms ($n$). When the number of Fourier terms is large, the size of the optimization vectors $A_n$ and $B_n$ also increases correspondingly. In the current work, we have set the number of Fourier waves to 100 ($n = 100$). The selection of the number of Fourier waves was made through a systematic search method to avoid both underfitting and overfitting. To determine the most suitable number, we evaluated the performance of various numbers of waves. Ten different values were tested: 1, 2, 5, 10, 20, 50, 100, 200, 500, and 1000. Through this testing, it was observed that the load profile was best fit using Fourier regression when the number of waves was set to 100. The demand quantiles (9 quantiles) for schools in the US using Fourier regression using historical data of one year are shown in **Fig. 2**. The demand quantiles are time series. Fourier regression for seven random consecutive days is shown in **Fig. 3**.

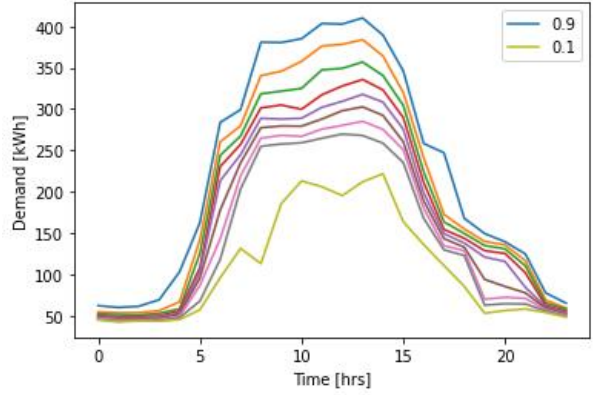

**Fig. 2** Electrical demand quantiles ($\beta = 0.1, 0.2, \ldots, 0.9$)

|     (a)     |     (b)     |
|-------------|-------------|

**Fig. 3** Fourier regression for one year of data, an example of seven random consecutive days with (a) $\beta = 0.5$, (b) $\beta = 0.7$

## 2.2. Probability Transition Matrix for Demand Quantiles

The corresponding demand quantile for every time step in the historical data of one year is extracted. Then it is used to calculate the probability of moving from one quantile to another. The probability matrix is shown in **Fig. 4**.

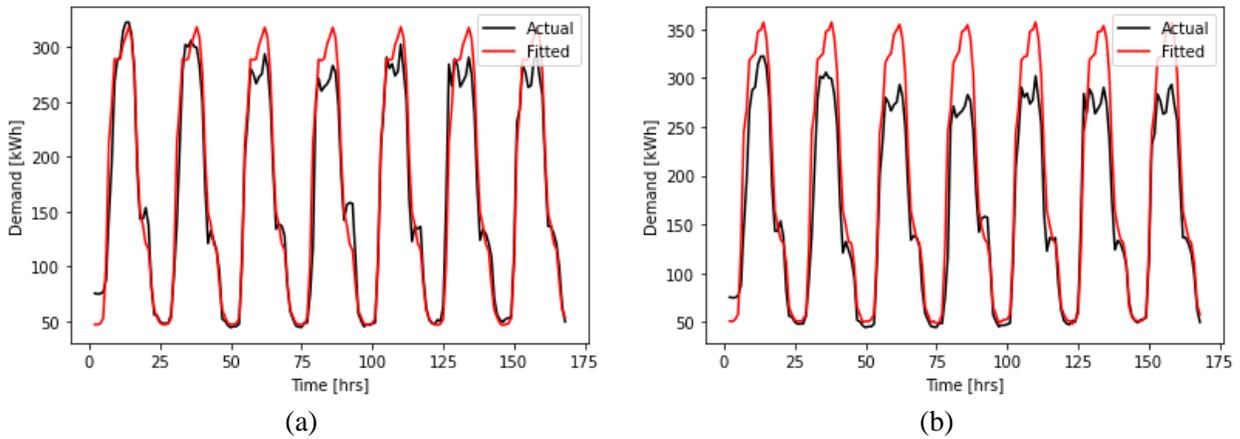

**Fig. 4** Transition matrix for demand quantiles

The label on the x and y axes are the demand quantiles. As is shown in **Fig. 4**, it is more likely to stay at the same demand quantile rather than moving to other demand quantiles. The probability of remaining at the same quantile ($\beta$) or moving to the neighboring quantiles ($\beta - 0.1$, $\beta + 0.1$) exceeds 70%. For instance, when starting at quantile $\beta=0.4$, there is a 31% chance of remaining in that quantile, a 21% probability of moving to quantile $\beta=0.3$, and a 19% probability of transitioning to quantile $\beta=0.5$. The combined probability of these three events amounts to 71%. It is less likely to move from quantile $\beta = 0.4$ to other quantiles. The trend holds for all demand quantiles. By utilizing Fourier quantile regression in this manner, the uncertainty of the demand is contracted.

## 2.3. MDP Formulation

Markov decision processes address the question of invoking the optimal action (amongst those available), in each state of the system. In general, an MDP aims to deliver an optimal decision rule, where the best action under each state is invoked. Optimality in this context is obtaining the minimum long run average cost (or equivalently maximum long run average profit). In the context of our problem, the state of our system is given by the time of day, state of charge of the battery and the quantile of demand for this period. The actions are charge and discharge of battery. The objective function in MDP is expressed as follows:

$$\text{minimize} \sum_{i=1}^{24} \sum_{j=0}^{10} \sum_{q=1}^{9} \sum_{k=1}^{21} y_{ijqk} \, C_{ijqk} \tag{4}$$

where $C_{ijqk}$ is the cost associated with each state; this will have been previously calculated. In this context, $y_{ijqk}$ represents the decision variable which signifies the probability of taking action $k$ and the system being in state $(i, j, q)$. Considering the system states $(i, j, q)$, $i$ is for the time of the day in an hour ($time = 1, 2, 3, \ldots, 24$), $j$ corresponds to the battery state of charge $S_{oc}$ ($S_{oc} = 0, 10, 20, \ldots, 100\%$), $q$ corresponds to the demand's value of a quantile $\beta$ ($\beta = 0.1, 0.2, 0.3, \ldots, 0.9$), and $k$ corresponds to the battery charging/discharging decision ($u = -1, -0.9, -0.8, -0.7, \ldots, 0, 0.1, \ldots, 1$).

The model is subject to the following constraints [23].

$$\sum_{i=1}^{24} \sum_{j=0}^{10} \sum_{q=1}^{9} \sum_{k=1}^{21} y_{ijqk} = 1 \tag{5}$$

$$y_{ijqk} \geq 0 \tag{6}$$

$$\sum_{k=1}^{21} y_{IJQk} - \sum_{i=1}^{24} \sum_{j=0}^{10} \sum_{q=1}^{9} \sum_{k=1}^{21} y_{ijqk} \, P_{iIjJqQ}(k) = 0$$
$$I = 1, 2, \ldots, 24, \tag{7}$$
$$J = 0, 1, \ldots, 10,$$
$$Q = 1, 2, \ldots, 9$$

Note that the above constraints reflect laws of probability for the Markov chain rendered through the MDP. This is a large-scale optimization problem:

- The dimension of $y_{ijqk}$ is 49896 ($24 \times 11 \times 9 \times 21$)
- The number of constraints is 52273 ($1 + 49896 + 2376$, $2376 = 24 \times 11 \times 9$)

The $P_{iIjJqQ}(k)$ is the probability of transition from state $(i,j,q)$ to state $(I,J,Q)$, which has been calculated beforehand as follows:

$$P_{iIjJqQ}(k) = P_{iI}(k) \times P_{jJ}(k) \times P_{qQ}(k) \tag{8}$$

In forming the above probability expression, we take care to note that $P_{iI}(k)$, the probability of moving from one time step to another, is independent of the decision $(k)$ and given as follows:

$$P_{iI}(k) = \begin{bmatrix} 0 & 1 & 0 & 0 & \cdots \\ 0 & 0 & 1 & 0 & \cdots \\ 0 & 0 & 0 & 1 & \cdots \\ \vdots & \vdots & \vdots & \vdots & \vdots \end{bmatrix} \tag{9}$$

$P_{jJ}(k)$ is dependent of the decision. For instance, $P_{jJ}(k)$ with $k = 9$ crossponding to $(u = -0.1)$ is given as follows:

$$P_{jJ}(k=9) = \begin{bmatrix} 1 & 0 & 0 & 0 & 0 & \cdots \\ 1 & 0 & 0 & 0 & 0 & \cdots \\ 0 & 1 & 0 & 0 & 0 & \cdots \\ 0 & 0 & 1 & 0 & 0 & \cdots \\ 0 & 0 & 0 & 1 & 0 & \cdots \\ 0 & 0 & 0 & 0 & 1 & \cdots \\ \vdots & \vdots & \vdots & \vdots & \vdots & \vdots \end{bmatrix} \tag{10}$$

For $k = 9$ $(u = -0.1)$, if the state of charge is 10% $(S_{oc} = 10\%, j = 1)$, the battery will be fully discharged $S_{oc} = 0$. If the state of charge is $S_{oc} = 0$ and $u = -0.1$, the battery can not discharge 10% of its maximum capacity as it does has any energy at this state. Thus, the battery will stay in the same state, $S_{oc} = 0$. Since the decision of $u = -0.1$ is not feasible at this state, a penalty is added to the cost function $C$, of equation (4) as shown in equation (16). $P_{jJ}(k)$ for other $k$ can be calculated in the same way.

In other words: In eq. (10), at state of $S_{oc} = 0$ (first row), if the descion is to discharge the battery by 10% $(u = -0.1)$, the energy stored in the battery will be below zero, which is not physically acceptable. However, by adding a penalty for violating these constraints, the MDP will avoid such decisions. In other words, $C$ includes the penalty on violatig the battery maximum/minimum capacity.

The probability transition matrix for demand quantiles $P_{qQ}(k)$ is given in **Section 2.2**, **Fig. 4**. $P_{qQ}(k)$ is extracted from histroical data and independent of the decision $k$.

## 2.4. Battery Scheduling Model

Total grid energy, $(E_G)$, is the amount of energy consumed by the modelled micro-grid (e.g. a school) from the power grid and is calculated as follows:

$$E_G(t) = E_L(t) + E_B(t) \tag{11}$$

where $E_L$ is the load of the facility which is uncontrollable, and $E_B$ is the battery energy given below [25]:

$$E_B(t) = C_B \max(\eta u(t), \frac{1}{\eta} u(t)) \tag{12}$$

$C_B$ is the battery's capacity, $\eta$ is the efficiency of charging or discharging (assumed equal), $u$ is the charging rate (positive for charging and negative for discharging).

The charging rate $u(t)$ is bounded as follows:

$$-1 \leq u(t) \leq 1 \tag{13}$$

The energy stored in the battery ($E_s$) at a given time ($t$) [25]::

$$E_s(t) = E_s(t-1) + C_B u(t) \tag{14}$$

The physical constraint on the energy stored in the battery is given:

$$0 \leq E_s(t) \leq C_B \tag{15}$$

The cost $C$, Eq. (4), is calculated as follows:

$$C_{ijqk} = E_{g\,ijqk} \times T + P \tag{16}$$

where $T$ is the electricity tariff, and $P$ ( $P = 1{,}000$) is the penalty for violating the physical constraint (15).

The state of charge ($S_{oc}$) is:

$$S_{oc}(t) = \frac{E_s(t)}{C_B} \tag{17}$$

## 2.5. MDP for Demand Peak Shaving

Demand peak shaving problem in its most straightforward format can not be addressed using MDP because the peak is a function of all previous states. In MDP, the expected current cost is independent of the previous states. However, in this section, the demand peak shaving problem is formulated to be an acceptable MDP chain.

The daily peak, i.e., the maximum hourly average demand recorded in a day, is calculated as shown below:

$$\text{Peak (n)} = \max\,[E_G\,(\text{t} = [n-1] \times N_i + 1: \text{n} \times N_i)] \tag{18}$$

$N_i$ is the number of steps in one day ($N_i = 24$, the day is 24 hours), and n is the day number within the year.

For the standard MDP to work, the immediate current expected cost must depend only on the current status and decision and so independent of the previous events. In eq. (18), the daily peak is dependent on all events that took place in the day. Therefore, the problem needs to be reformulated to address this dependency.

In the new formulation, the peak will be represented as a vector, a day in length, recording the maximum demand that occurred so far along the day. The state of the peak is given as shown below:

$$Peak(t+1) = \max\,[Peak(t), E_G(t)\,] \tag{19}$$

The $Peak$ will be considered as one of the system's MDP states (other states are the time of the day ($t$), the demand quantile ($\beta$), and the battery state of charge ($S_{oc}$).

The peak vector is reset at the start of each day, with an initial state of zero for the peak at the beginning of the day.

$$Peak(t = 0) = 0 \tag{20}$$

The value of the peak is divided into 6 thresholds, as shown in **Table 1**.

**Table 1**. Peak thresholds

| $r$ | $Peak\ (r)$ |
|---|---|
| 1 | $Peak \leq 100$ |
| 2 | $Peak \geq 100$ |
| 3 | $Peak \geq 200$ |
| 4 | $Peak \geq 300$ |
| 5 | $Peak \geq 400$ |
| 6 | $Peak \geq 500$ |

MDP equations:

The objective function is:

$$\text{minimize} \sum_{i=1}^{24} \sum_{j=0}^{10} \sum_{q=1}^{9} \sum_{r=1}^{6} \sum_{k=1}^{21} y_{ijqrk}\ C_{ijqrk} \tag{21}$$

The model is subject to the following constraints [23]:

$$\sum_{i=1}^{24} \sum_{j=0}^{10} \sum_{q=1}^{9} \sum_{r=1}^{6} \sum_{k=1}^{21} y_{ijqrk} = 1 \tag{22}$$

$$y_{ijqrk} \geq 0 \tag{23}$$

$$\sum_{k=1}^{21} y_{IJQRk} - \sum_{i=1}^{24} \sum_{j=0}^{10} \sum_{q=1}^{9} \sum_{r=1}^{6} \sum_{k=1}^{21} y_{ijqok}\ P_{iIjJqQrR}(k) = 0$$
$$I = 1,2, \dots, 24,$$
$$J = 0, 1, \dots, 10,$$
$$Q = 1,2, \dots, 9$$
$$R = 1,2, \dots, 6$$
$$\tag{24}$$

This is another large-scale optimization problem:

- The dimension of $y_{ijqrk}$ is 299376 ($24 \times 11 \times 9 \times 6 \times 21$)

- The number of constraints is 313633 ($1 + 299376 + 14256,\ 14256 = 24 \times 11 \times 9 \times 6$)

where $P_{iIjJqQrR}$ is given below:

$$P_{iIjJqQrR}(k) = P_{iI}(k) \times P_{jJ}(k) \times P_{qQ}(k)\ P_{rR}(i, q, k) \tag{25}$$

The probability of moving from one peak threshold to another threshold $P_{rR}(i, q, k)$ is dependent on the current demand value and the current decision (index $k$). The current demand value is dependent of the current demand quantile (index $q$) and current time (index $i$). The peak value is reset to zero at the beginning of the day.

To calculate the transition matrix $P_{rR}$ for system states $(i, q)$ and decision $k$, we use equation (19) to find the peak threshold the system will move to.

## 3. Results and Discussions

To reduce the complexity of the current optimization problem, the equality constraint was relaxed in **Appendix A.1**. While the policy generated by MDP sometimes violates the physical constraints, the policy is corrected in **Appendix A.2**. The performance of MDP is gauged using the ideal dispatching decision that is given in **Appendix A.3**. A comprehensive case study is conducted to validate the proposed model, the details of which are provided in Section 3.1. The validation results for different pricing policies are shown in **Section 3.2**.. The computing resources utilized are described in **Section 3.3**.

### 3.1. Case Study

An electrical load for a school in the USA is used to conduct this study. Historical data of two years is utilized to validate MDP in scheduling battery systems under load uncertainty at different electricity pricing policies. One year is used to extract the demand quantiles, the transition probability matrices, and MDP policy. The remaining year of data (unseen) is used to validate the MDP policy. The relevant parameters used are summarized in **Table 2**.

**Table 2.** Parameters

| Parameters | Value |
|---|---|
| Battery capacity, $C_B$ | 500 [$kWh$] |
| Battery initial state of charge, $S_{oc}|_{t=0}$ | 1 |
| Charging/discharging efficiency, $\eta$ | 0.92 |

Four pricing policies are used to test MDP. These pricing policies are summarized in **Table 3**. Pricing policies A and B are shown in the following figure, **Fig. 5**.

**Table 3.** Pricing Policies

| Policy | Electricity tariff | Scheduling complexity under load uncertainty |
|---|---|---|
| A | Time-of-use [27]: off-peak, peak, shoulder | Intuitive |
| B | Real-time [28] | Moderate |
| C | Energy-limit penalty:<br>$E_G(t) > 200$[kWh], price = \$0.12/kWh<br>$E_G(t) \leq 200$[kWh], price = \$0.05/kWh | Complex |
| D | Peak demand disincentives [29]:<br>price = \$0.05/kWh, \$7/kW peak | Very complex |

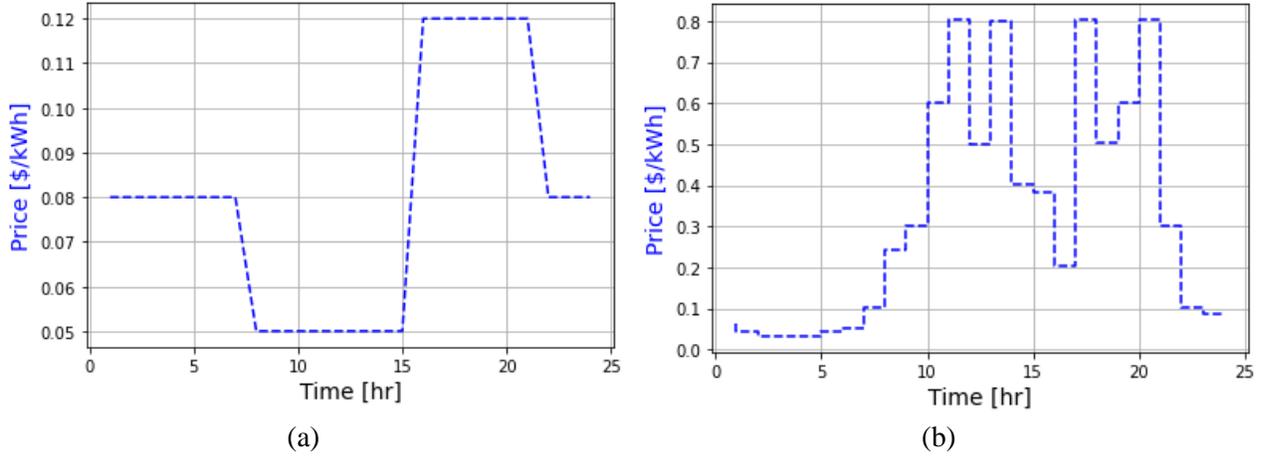

**Fig. 5** Pricing policies (a) Policy A [27] (b) Policy B [28]

### 3.2. MDP Testing and Validation

MDP is trained using historical data from one year and then tested on another year's data. The performance is gauged by comparing the bill amount obtained by implementing MDP policy against the ideal bill amount. The ideal bill amount represents the optimal scenario where scheduling decisions are made with perfect knowledge of demand over the year, eliminating any load uncertainty. The results for different pricing policies over one year are summarized in **Table 4**. Each pricing policy (A-D) is evaluated under three cases. In case 1, the battery charging and discharging losses are ignored, and the customers are allowed to sell energy to the grid. In case 2, the battery charging and discharging losses are taken into account in the decision-making process, while customers are still permitted to sell energy to the grid. In case 3, the battery charging and discharging losses are included, but the customers will not be refunded for supplying energy to the grid.

MDP is reliable for pricing policies A and B, with an efficiency of at least 90% in all cases. However, scheduling the battery under pricing policy C proves to be more complex (**Table 3**); MDP achieves efficiency in the 67-78% range. MDP efficiency in scheduling battery systems for shaving the demand peak is only 10%. To enhance the MDP's efficiency in shaving the demand peak, the number of peak thresholds is increased to 11 and 21. The new MDP's efficiency in shaving the demand peak is shown in **Table 5**.

**Table 4.** Results summary of MDP performance

| Pricing | Bill with no battery [$] | Bill with battery [$] | | Saving[$] | | MDP efficiency [%] |
|---|---|---|---|---|---|---|
| Policy A, Case #: | | MDP | Ideal | MDP | Ideal | |
| 1 | 68,245 | 59,279 | 59,280 | 8964 | 8965 | 99.98 |
| 2 | 68,245 | 61071 | 61062 | 7173 | 7183 | 99.86 |
| 3 | 68,245 | 61,939 | 61,245 | 6,306 | 7000 | 90.08 |
| Policy B, Case #: | | MDP | Ideal | MDP | Ideal | |
| 1 | 37,608 | 12,501 | 12,499 | 25,107 | 25,109 | 99.99 |
| 2 | 37,608 | 18,753 | 17,148 | 18,855 | 20,460 | 92.15 |
| 3 | 37,608 | 25,212 | 23,803 | 12,396 | 13,806 | 89.78 |
| Policy C, Case #: | | MDP | Ideal | MDP | Ideal | |
| 1 | 88,361 | 67,714 | 57,707 | 20,647 | 30,654 | 67.35 |
| 2 | 88,361 | 69454 | 60,633 | 18906 | 27,728 | 68.18 |
| 3 | 88,361 | 66,723 | 60,633 | 21,637 | 27,728 | 78.03 |
| Policy D, Case #: | | MDP | Ideal | MDP | Ideal | |
| 1 | 574,832 | 561,158 | 437,666 | 13674 | 137166 | 9.97 |
| 2 | 574,832 | 562,102 | 438,981 | 12730 | 135851 | 9.37 |
| 3 | 574,832 | 561,158 | 437,666 | 13674 | 137166 | 10 |

**Table 5**. MDP efficiency in shaving the demand peak with different peak thresholds (different increments)

| Peak Thresholds Count | Peak Increment [kW] | MDP Efficiency [%] | Optimization Variables Count | Constraints Count | Computation Time |
|---|---|---|---|---|---|
| 6 | 100 | 10 | 272,160 | 285,151 | 9 hrs |
| 11 | 50 | 30.8 | 498,960 | 522,721 | 2 days |
| 21 | 25 | 29.1 | 952,560 | 997,921 | 5 days |

Increasing the number of peak thresholds from 6 to 11 increases MDP efficiency. However, contrary to expectations, further increasing the number of peak thresholds from 11 to 21 does not enhance MDP efficiency. Increasing the number of peak thresholds increases the number of optimization variables and constraints. The optimization problem with 21 thresholds has about 950 thousand optimization variables and nearly 1 million constraints. This numerical burden poses challenges for solving the problem effectively. As a result, when utilizing the Gurobi solver, it returns a solution that violates the previously relaxed equality constraint discussed in **Appendix A.1**. The maximum violation was about 0.00815.

**Fig. 6** - **Fig. 9** show battery scheduling examples for a typical day for different pricing policies. In **Fig. 6** (a), the battery is empty ($E_S = 0$) at 0:00. Between 7:00 and 14:00, when the price of electricity is at the lowest level, the battery is set to charge mode at 7:00. To achieve the best energy cost reduction, the battery is fully charged by 14:00 and then is set to discharge mode after then, reaching full discharge

by 20:00. The battery is better to be left fully charged till 14:00 and then fully discharge it during the high price period between 14:00 and 20:00-. Discharging the battery between 8:00 and 10:00, as shown in **Fig. 6** (a), is not reducing the cost of electricity. Recharging the battery between 10 and 12 corrected this decision. Without penalty on charging/discharging, the scheduling problem has more than one optimal solution that explains the observed oscillations in the battery schedule. Introducing a penalty on charging/discharging will not only be a more accurate representation of the battery operation but also produces a smooth policy, as in **Fig. 6** (b), that guarantees that the charging/discharging process contributes less to the battery aging.

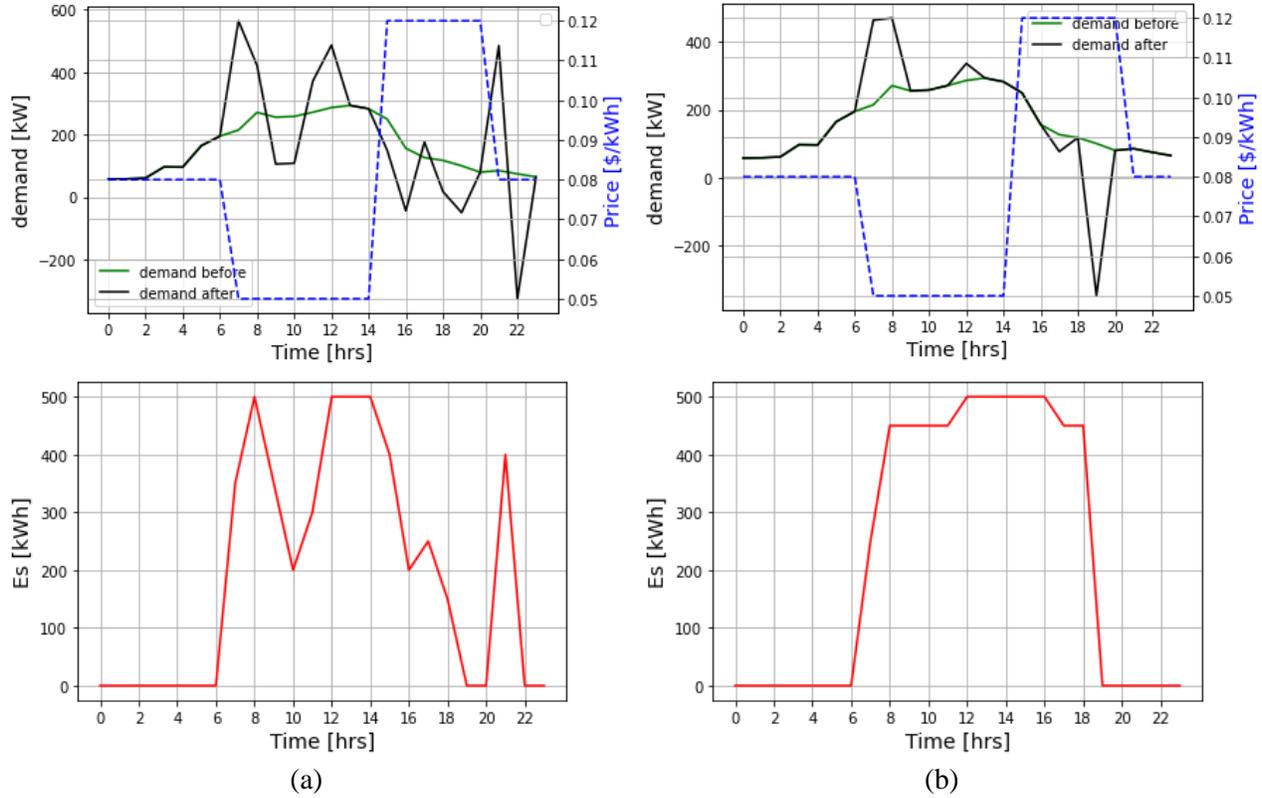

(a)          (b)

**Fig. 6** Demand and battery energy profiles for one random day with Policy A (a) penalty on charging/discharging is ignored (Case #1) (b) penalty is included (Case #2)

Pricing policy B is more challenging than pricing policy A in terms of load uncertainty's effect on the battery schedule's optimality due to the oscillating electricity prices. However, MDP policy is found to be reliable in handling load uncertainty. In **Fig. 7,** we see how MDP policy avoids high electricity consumption during the high-price period for two different days.

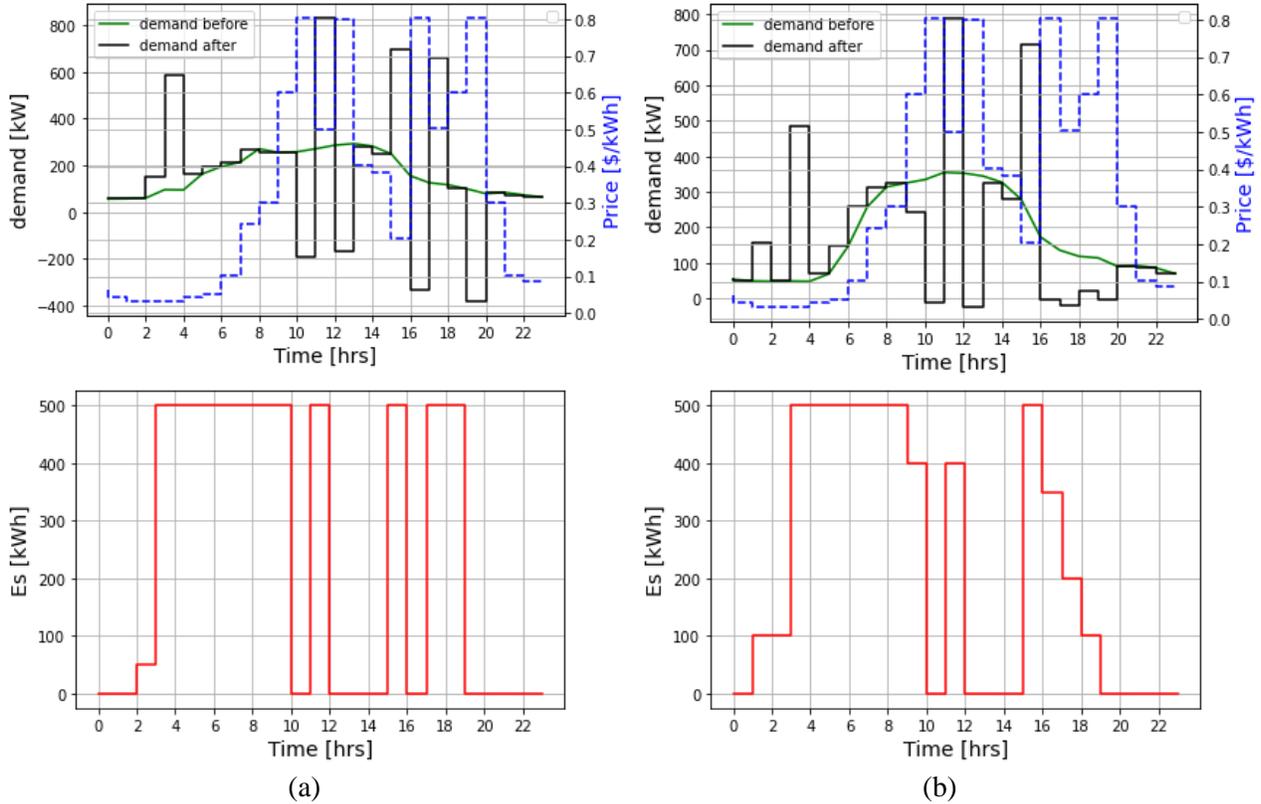

(a)                            (b)

**Fig. 7** Demand and battery energy profiles for one random day with policy B (a) typical day (b) another typical day

Scheduling battery systems with MDP under pricing policy C is shown in **Fig. 8** for a typical day. The battery is in charging mode between 0:00 and 3:00 with a charging rate that keeps the demand energy below 200 kWh. The battery is switched to discharge mode between 06:00 and 11:00 to maintain the demand below 200 kWh to avoid the high prices. The battery is fully discharged by 11:00. The battery is then set to charging mode and fully charged by 12:00. The battery is then fully discharged once again.

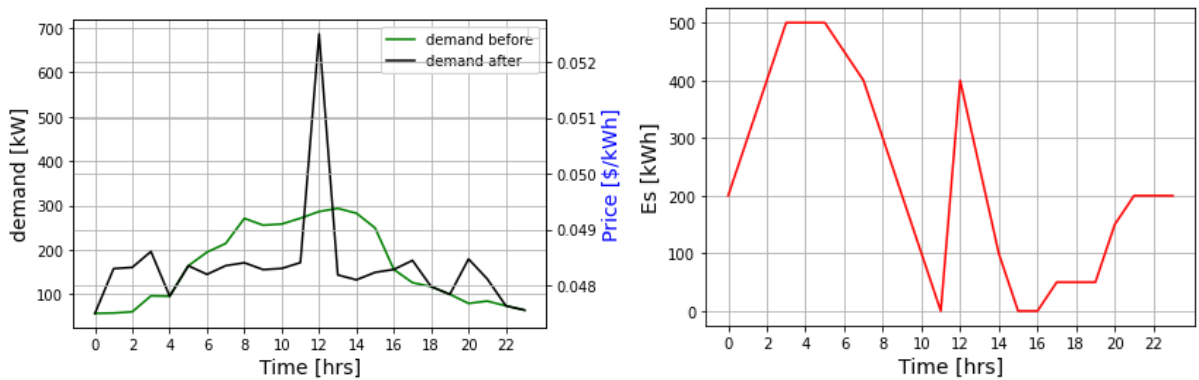

**Fig. 8** Demand and battery energy profiles for one random day with policy C for a typical day

Shaving the demand peak (policy D) under load uncertainty proves to be the most challenging. MDP recorded an efficiency of only 30.8%, as in **Table 5**. To better understand the performance, the demand profile with MDP can be compared with the ideal one in **Fig. 9**. In **Fig. 9**, shaving the demand peak under load uncertainty is illustrated for one random day using (a) MDP and (b) the scenario where the optimal decision is applied. MDP shaves the demand peak from 340 to 290 kWh. If the battery schedule is ideal, the demand is shaved from 340 kWh to 245 kWh. The performance of MDP in shaving the demand peak of this day is %52.6. MDP, when tested on one year in which different energy consumption patterns are observed, achieves an efficiency of only 30.8%.

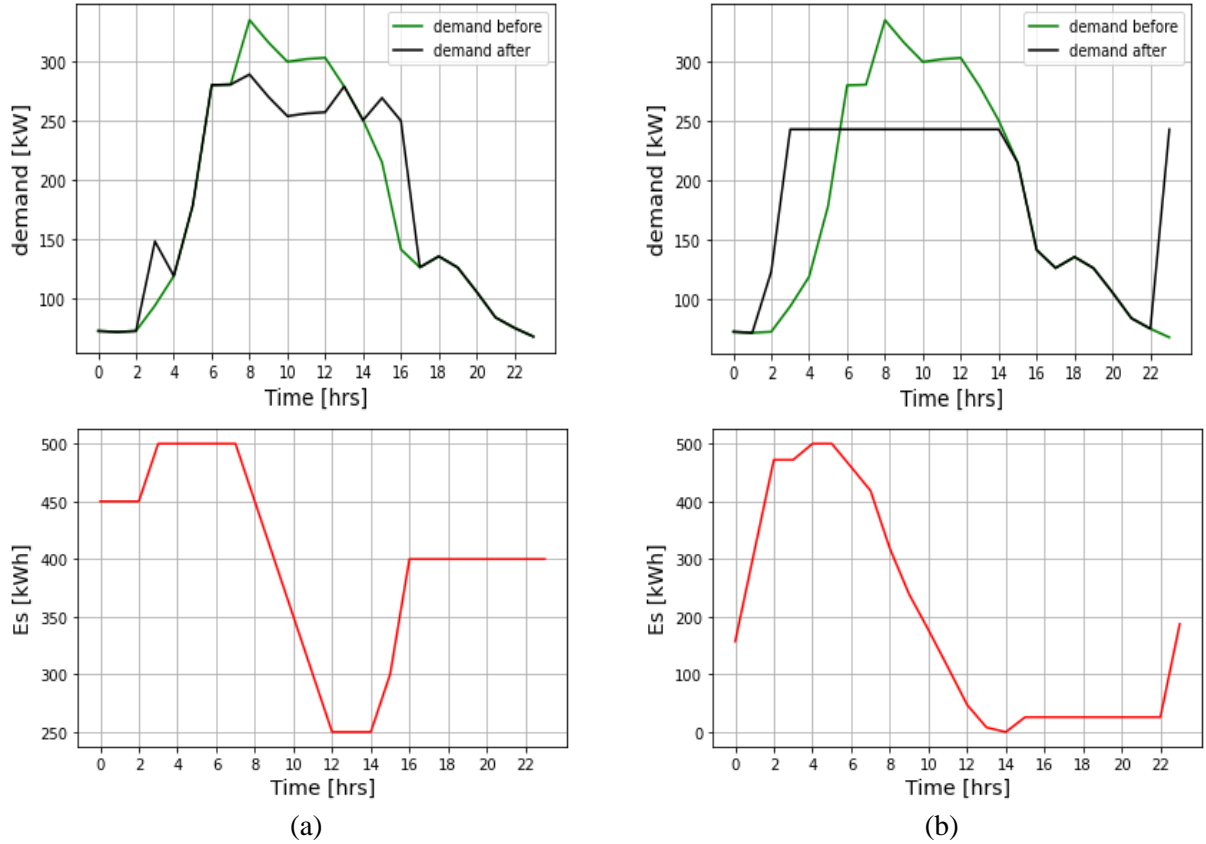

(a)　　　　　　　　　　　　　　　(b)

**Fig. 9** Demand and battery energy profiles for one random day with policy D (a) MDP (b) Ideal schedule

### 3.3. Codes Repository and Computation Resources

The Python codes, used to conduct this research, are publicly available on GitHub [24]. The optimization process involved the utilization of the Gurobi solver, which typically required 8-20 minutes to compute the optimal MDP policy. However, for peak shaving problems with 6 peak thresholds, it consumes about 9 hours due to the large number of optimization variables and constraints. It takes more if more thresholds are assumed. The simulations were executed on Lenovo X1 Carbon business laptop with Intel(R) Core(TM) i7-8665U CPU @ 1.90GHz　2.11 GHz and 16.00 GB RAM.

### 4. Conclusion

In this study, a real-time Markov decision process (MDP) was used to construct hourly battery schedule to reduce the electricity cost for buildings under load uncertainty. MDP was based on simulating the uncertainty by fitting the load using Fourier quantile regression and calculating the quantile probability of transition. To assess the MDP's effectiveness, MDP was tested in scheduling the battery systems over the course of one year. The performance of MDP in scheduling the battery systems under load uncertainty was found to be affected by the complexity of the pricing policy. The performance of Markov decision process can vary significantly, ranging from 30 to 99%. This study considers various pricing policies, among which demand peak shaving stands out as the most complex and challenging case.

**5.6 Declaration of generative AI and AI-assisted technologies in the writing process**

During the preparation of this work the authors used ChatGPT in order to rephrase some statements to enhance the readbility. After using this tool, the authors reviewed and edited the content as needed and take full responsibility for the content of the publication.

28. Rajamand, S.: Effect of demand response program of loads in cost optimization of microgrid considering uncertain parameters in PV/WT, market price and load demand. Energy. 194, 116917 (2020). https://doi.org/10.1016/j.energy.2020.116917

29. Industrial Assessment Centers, https://www.energy.gov/eere/amo/industrial-assessment-centers-iacs

30. Sharadga, H., Hajimirza, S., Cari, E.P.T.: A Fast and Accurate Single-Diode Model for Photovoltaic Design. IEEE J. Emerg. Sel. Top. Power Electron. (2020). https://doi.org/10.1109/jestpe.2020.3016635

**Appendix**

**Appendix A.1. Equality Constraints Relaxation**

Solving an optimization problem with an extensive set of equality constraints and a large number of optimization variables of a small scale can be a formidable task. To tackle this challenge, the equality constraints are relaxed, as outlined in the following table (**Table 6**). In the current work, relaxation # 3 is adopted to extract the results. The maximum violation for this case was negligible and can be disregarded. For choice 1, the model remains infeasible even after scaling the variables; the solver fails to respect the equality constraints due to the complexity of solving for a huge number of equality constraints.

**Table 6.** Equality constraints relaxation options for policy C

| # | Equations (5), (7): left side | Best objective | Solution | Non-violation check of constraint | |
|---|---|---|---|---|---|
| | | | | **Equation (5)** | **Equation (7)** |
| 1 | '= 1', '= 0' | - | Infeasible | - | - |
| 2 | '= 1', '$\geq 0$' | - | Infeasible | - | - |
| 3 | '= 1', '$\leq 0$' | 10.036 | Deterministic policy | Pass | Pass violation $< 1.6 \times 10^{-20}$ |
| 4 | '$\leq 1$', '$\leq 0$' | 0 | Zeros; not acceptable | Fails | Pass |
| 5 | '$\geq 1$', '$\leq 0$' | 10.036 | Deterministic policy | Pass | Pass violation $< 3.1 \times 10^{-18}$ |
| 6 | '= 1', '$\leq 0$' & '$\geq -0.0001$' | 10.058 | Probabilistic policy for 11 out of 2,376 states ($24 \times 11 \times 9$) | Pass | Pass violation $< 3.4 \times 10^{-18}$ |

**Appendix A.2. MDP Policy Corrections**

The decision obtained by MDP sometime violates the constraints on the energy stored in the battery, which are given in eq. (15). The policy is corrected as follows to satisfy these constraints:

$$\begin{aligned} if\ E_s(t) > C_B \quad & E_s(t) = C_B\ \&\ u(t) = 1 - S_{oc}(t-1) \\ if\ E_s(t) < 0 \quad & E_s(t) = 0\ \&\ u(t) = -S_{oc}(t-1) \end{aligned} \quad (26)$$

For some states, MDP returns no policy, i.e., indicating that it is unlikely for the system to be in these states. An example of these states is when the battery is fully charged during the peak period. To handle these states in case of occurrence, the policy for these states is set to refrain from both charging and discharging actions, $u = 0$. For some states, MDP returns multiple MDP policies, resulting in a probabilistic policy. To handle this situation, the probabilities of the policies associated with these states are stored.. The Monte Carlo concept is then employed to randomly select a policy based on these probabilities.

**Appendix A.3. Ideal Decision**

In the ideal decision problem, the load for the year is assumed to be previously given. Utilizing this information, the battery schedule is optimized, and the resulted bill amount is used to calculate how efficient MDP is to make a decision under the uncertainty of the electrical load.

In the ideal case, the charging rate vector ($u(t)$) is optimized over one year., resulting in a vector of length 8760 ($24 \times 365$). However, optimizing such large vector is computationally expensive. To address this challenge and reduce computation time, the optimization problem is reformulated in this work by utilizing convex optimization or linear programming. Solving the scheduling problem with pricing policy B is given in detail in this paper. Formulating the scheduling problem for other pricing policies is illustrated in the Python codes, which were made available on GitHub [24].

The cost ($C$) is given as follows:

$$C(t) = 0.12 \times E_G(t), \ if \ E_G(t) > 200$$
$$C(t) = 0.05 \times E_G(t), \ if \ E_G(t) \leq 200 \tag{27}$$

To convert the "if" statement into a linear program, we introduce a binary variable $B$ as shown below [25]:

$$C(t) = 0.05 \times E_G(t) + 0.07 \times B(t) \times E_G(t) \tag{28}$$

$B$ is 1 if $E_G(t) > 200$ and zero if $E_G(t) \leq 200$. That can be rewritten using the big-M method by introducing the following two constraints [25]::

$$E_G(t) > 200 - M\,(1 - B(t))$$
$$E_G(t) \leq 200 + M\,B(t) \tag{29}$$

In eq. (28), $B$ and $E_G$ are optimizaion variables. The product of two optimization variables produces a bilinear problem. To linearize the product of binary variable $B$ and continous varsible $E_G$, we replace the product with a new continuous variable $Z$ and add new four constraints as instructed in [26]. However, the continuous variable must be bounded. To solve this issue, we replace $E_G$ using eq. (11).

$$C(t) = 0.05 \times E_G(t) + 0.07 \times B(t)\,[E_L(t) + E_B(t)]$$
$$C(t) = 0.05 \times E_G(t) + 0.07 \times B(t) \times E_L(t) + 0.07 \times B(t)\,E_B(t) \tag{30}$$

The bilinear term is $B(t)\,E_B(t)$, $E_B(t)$ is bounded as shown below:

$$-C_B \times \eta \leq E_B(t) \leq \frac{C_B}{\eta} \tag{31}$$

We linearize eq. (30) using new varaible ($Z$):

$$C(t) = 0.05 \times E_G(t) + 0.07 \times B(t) \times E_L(t) + 0.07 \times Z(t) \tag{32}$$

For this formulation to hold, we add four new constraints as instructed in [26]. The four constraints are given as follows:

$$-C_B\,\eta\,B \leq Z(t) \leq \frac{C_B}{\eta}\,B \tag{33}$$

$$-C_B\,\eta\,(1-B) \leq E_B(t) - Z(t) \leq \frac{C_B}{\eta}\,(1-B)$$

Gurobi solver is used to solve all optimization problems in the current work. However, the presence of max function in eq. (12) requires reformulation to make it compatible with Gurobi solver. That can be done by introducing new auxiliary optimization variables and max constraint. The max constraint is nonconvex, but Gurobi solver handles it by converting it into a MIP inequality constraint. Alternatively, a more efficient way to formulate the max function can be achieved without using binary variables by introducing the following two constraints [25]::

$$E_B(t) \geq C_B\,\eta\,u(t)$$

$$E_B(t) \geq C_B\,\frac{1}{\eta}\,u(t) \tag{34}$$

Since we minimize the cost ($C$), which is given in eq. (30), the battery energy ($E_B$) need to be minimized. The minimum possible value for $E_B$ is either $C_B\,\eta\,u(t)$ or $C_B\,\frac{1}{\eta}\,u(t)$ according to eq. (34). To satisfy the two constraints of eq. (34), $E_B$ must be equal to the larger of the two values ($C_B\,\eta\,u(t)$, $C_B\,\frac{1}{\eta}\,u(t)$). Thus these two constraints serve as substitue for the max function.

The new optimization problem after reformulation is given below:

$$\text{Minimize } C(t) = 0.05 \times E_G(t) + 0.07 \times B(t) \times E_L(t) + 0.07 \times Z(t) \tag{35}$$

Subject to the following constraints:

$$E_G(t) > 200 - M\,(1 - B(t)) \tag{36}$$

$$E_G(t) \leq 200 + M\,B(t)$$

$$-C_B\,\eta\,B \leq Z(t) \leq \frac{C_B}{\eta}\,B$$

$$-C_B\,\eta\,(1-B) \leq E_B(t) - Z(t) \leq \frac{C_B}{\eta}\,(1-B)$$

$$E_B(t) \geq C_B\,\eta\,u(t)$$

$$E_B(t) \geq C_B\,\frac{1}{\eta}\,u(t)$$

$$-1 \leq u(t) \leq 1$$

Gurobi solver consumes about 12 minutes to return the solution of the ideal decision problem. The complexity of the problem arises from the large number of binary variables involved, as the vector $B(t)$ has alength of 8760.